**Detection of magnetized quark-nuggets, a candidate for dark matter.**

J. Pace VanDevender[1]*, Aaron P. VanDevender[2], T. Sloan[3], Criss Swaim[4], Peter Wilson[5], Robert. G. Schmitt[6], Rinat Zakirov[1], Josh Blum[1], James L. Cross Sr.[7], and Niall McGinley[8]

[1]VanDevender Enterprises LLC, 7604 Lamplighter LN NE, Albuquerque, NM 87109 USA. [2]Founders Fund, One Letterman Drive, Building D, 5th Floor, Presidio of San Francisco, San Francisco, CA 94129 USA. [3]Department of Physics, Lancaster University, Lancaster, LA1 4YB, UK. [4]The Pineridge Group, 2458 Mapleton Ave., Boulder, CO 80304 USA. [5]School of Geography and Environmental Sciences, Ulster University, Cromore Road, Coleraine, Co. Londonderry, BT52 1SA Northern Ireland, UK. [6] Sandia National Laboratories, Albuquerque, NM 87185-0840. [7]Cross Marine Projects,1021 Pacific Ave., American Fork, UT 84003. [8]Ardaturr, Churchill PO, Letterkenny, Co. Donegal, Ireland.

*pace@vandevender.com

Quark nuggets are theoretical objects composed of approximately equal numbers of up, down, and strange quarks and are also called strangelets and nuclearites. They have been proposed as a candidate for dark matter, which constitutes ~85% of the universe's mass and which has been a mystery for decades. Previous efforts to detect quark nuggets assumed that the nuclear-density core interacts directly with the surrounding matter so the stopping power is minimal. Tatsumi found that quark nuggets could well exist as a ferromagnetic liquid with a ~$10^{12}$-T magnetic field. We find that the magnetic field produces a magnetopause with surrounding plasma, as the earth's magnetic field produces a magnetopause with the solar wind, and substantially increases their energy deposition rate in matter. We use the magnetopause model to compute the energy deposition as a function of quark-nugget mass and to analyze testing the quark-nugget hypothesis for dark matter by observations in air, water, and land. We conclude the water option is most promising.

## Introduction

About 85%[1] of the universe's mass[2-4] does not interact strongly with light; it is called dark matter[5]. Extensive searches for a subatomic particle that would be consistent with dark matter have yet to detect anything above background signals[6-12]. Macroscopic quark nuggets[13], which are also called strangelets[14] and nuclearites[15] are theoretically predicted objects[13-15] composed of three of the six types of quarks. Quarks are the basic building blocks of protons, neutrons, and many other particles in the Standard Model[16]. Quark nuggets[13-14, 17] interact[5, 17-19] with all matter through the gravitational force and with each other through the strong nuclear force. There have been many searches[20] for nuclearites with underground detectors which would not be sensitive to highly magnetic ones, as proposed by Tatsumi[21], because of the greatly enhanced stopping power of the magnetized quark nuggets. In this paper, we consider such highly magnetic quark nuggets.

Detecting them is a challenge. The original theory for quark nuggets by Witten[13] shows their density should be somewhat larger than the density of nuclei and their mass can be very large. The large mass means the number per unit volume of space is small, so one would rarely hit the earth and would require a very large detector to detect them. The gravitational force on even a 0.001 kg quark-nugget produces ~ 10 million atmospheres pressure on adjacent material below it, so any quark nugget that has been stopped in the earth will sink to the center of the earth.

However, they can be detected by their interaction with matter as they are slowed down from astrophysical velocity of ~ 250 km/s.

Full Quantum Chromo Dynamics (QCD) calculations of quark-nugget formation and stability are still impractical, so the successful MIT Bag Model has been used with its inherent limitations.[22] Large quark nuggets are predicted to have formed and be stable[13, 14, 23, 24] with mass between $10^{-8}$ kg and $10^{20}$ kg[23] within a plausible but uncertain range of assumed parameters of QCD and the Bag Model, assuming a first-order phase transition.

None of these studies considered the effect of the intense magnetic field on stability. Chakrabarty[25] showed that the stability of quark nuggets increases with increasing magnetic field for $B \leq 10^{16}$ T, so the large self-field described by Tatsumi[21] should enhance their stability.

Aoki, *et al.*[26] showed that this finite-temperature QCD transition in the hot early universe was very likely to have been an analytic crossover, involving a rapid change as the temperature varied, but a not a real phase transition. However, Atreya, *et al*.[27] have found another mechanism that should form quark and anti-quark nuggets, regardless of the order of the quark-hadron phase transition; their mechanism is based on CP-violating quark and anti-quark scatterings from moving Z(3) domain walls. Recent simulations by T. Bhattacharya, *et al*.[28] support the crossover process. Experiments by A. Bazavov, *et al*.[29] at the Relativistic Heavy Ion Collider (RHIC) have provided the first indirect evidence of strange baryonic matter. Additional experiments at RHIC may determine whether the process is a first order phase transition or the crossover process. In either case, quark nuggets are theoretically viable forms of matter.

In 2001, Wandelt, *et al.* [17] showed that quark nuggets meet all the theoretical requirements for dark matter and is not excluded by observations when the stopping power for quark nuggets in the materials covering a detector is properly considered and when the average mass is > $10^5$ GeV ($\sim 2 \times 10^{-22}$ kg). In 2014, Tulin[19] surveyed additional simulations of increasing sophistication and updated the results of Wandelt, *et al*. The combined results help establish the allowed range and velocity dependence of the strength parameter and strengthen the case for quark nuggets. In 2015, Burdin, *et al.*[30] examined all non-accelerator candidates for stable dark matter and also concluded that quark nuggets meet the requirements for dark matter and have not been excluded experimentally.

These results are not generally known, and there seems to be a general understanding that quark nuggets have been excluded by observations. For example, the paper by Gorham and Rotter[31] about constraints on anti-quark nugget dark matter (which do not constrain quark-nuggets unless the ratio of anti-quark nuggets to quark nuggets is shown to be large) assumes that limits on the flux of magnetic monopoles from analysis by Price, *et al.*[32] of geologic mica buried under 3 km of rock are also applicable to quark nuggets. Their conclusion requires that the cross section for momentum transfer be simply the size of the nuclear-density core.

Gorham and Rotter also cite work by Porter, *et al.*[33-34] as constraining quark-nugget contributions to dark matter by the absence of meteor-like objects that are fast enough to be quark nuggets. Porter, *et al.*'s analysis was based on the early work by De Růjula and Glashow[15], who also assumed the geometric cross section of the quark nugget is appropriate for momentum transfer.

Witten's value for the mass density is "somewhat greater than nuclear mass density" or > $2.1 \times 10^{17}$ kg/m$^3$ and indicates quark-nuggets are in the theoretically predicted, ultra-dense, Color

Flavor Locked (CFL) phase[24] of quark matter. Steiner, *et al.*[35] showed that the ground state of the CFL phase is color neutral and that color neutrality forces electric charge neutrality, which minimizes electromagnetic emissions. So quark nuggets are both dark and very difficult to detect with astrophysical observations.

The large density does indeed give a very small geometrical cross section. As we will explain, the effective cross section is greatly enhanced if they are magnetized. Tatsumi[21] calculates the value of the magnetic field at the surface of a quark-nugget core inside a neutron star to be $10^{12\pm1}$ T, which is large compared to expected values for the magnetic field at the surface of a neutron star. For a quark nugget of radius $r_{qn}$ and a neutron star or radius $R$, Tatsumi[21] finds that the magnetic field scales as $(r_{qn}/R)^3$. Therefore, the surface magnetic field of a neutron star is substantially smaller than $10^{12}$ T because $R > r_{qn}$. However, quark-nugget dark matter is bare, i.e. $(r_{qn}/R) = 1$, so the surface magnetic field of what we wish to detect is $10^{12\pm1}$ T. In this paper, we explore the consequences of that large magnetic field on the detectability of quark nuggets.

We have estimated the flux of electromagnetic radiation from electrons that might be swept up by the magnetic field of quark-nuggets in gravitational motion through the galaxy and by their synchrotron radiation from a magnetized quark nugget interaction with the Galactic magnetic field. Both proved to emit an energy flux which is orders of magnitude less than the cosmic microwave background.

Although we will show that the cross section for interacting with dense matter is greatly enhanced by the magnetic field, the field falls off as radius $R^{-3}$ and the collision cross section is still many orders of magnitude too small to violate the collision requirements[17, 19, 30] for dark matter. In addition, electric-charge neutral, magnetized quark nuggets do not emit electromagnetic radiation unless they are rotating. Even if some process caused it to be formed with a high rotational velocity, it would quickly lose its rotational energy. For example, using Eq. 2 from Pacini[36], the radiated power for a 1-kg quark nugget (with mass density of $10^{18}$ kg/m$^3$) rotating at $10^6$ Hz is $P = 1.7 \times 10^{-8}$ W. Since the rotational kinetic energy $W_r = 3.55$ J, the slowing down time is $W_r/P = 2 \times 10^8$ s = 6.6 y. Therefore, on cosmological times, the quark nuggets would be brought to rest rotationally and would not radiate significantly. They would not emit observable electromagnetic waves today. We conclude that QNs in the mass range considered here are unlikely to emit detectable radiation and, therefore, are candidates to contribute to the dark matter.

The theories for the existence, stability, and properties of quark nuggets are still evolving. For example, the work by Witten[13] and Fahri[14] use the MIT Bag Model with its limitations[22] because full QCD calculation are still not practical. Consequently, we see the proposed detection of quark-nuggets as high-risk; they may not exist. However, if they are detected, the results should motivate more powerful theories and better observations and experiments.

### Enhanced energy deposition by magnetopause effect

The cross section $Q_n$ for direct interaction of a quark nugget of mass $M$ with the surrounding matter has been assumed to be the cross sectional area of its core of mass density $\rho_{qn} \sim 2.1 \times 10^{17}$ kg/m$^3$

$$Q_n = \pi\left(\frac{3M}{4\pi\rho_{qn}}\right)^{\frac{2}{3}} \qquad (1)$$

However, ionized matter flowing past the magnetized quark nugget (in the rest frame of the quark nugget) is similar to the solar wind passing the earth, as illustrated in Fig. 1.

Since the earth's magnetosphere and ionosphere consists of ionized matter (plasma), the magnetosphere model is appropriate at those high altitudes. As the quark nugget proceeds into the troposphere and finally into water or soil, it encounters un-ionized (electrically neutral) material. However, we assume the material is ionized by the quark-nugget plasma, compute the plasma properties at each point in its trajectory, and then compute the ionization rate of neutrals flowing into that plasma to test the assumption about ionization. We find that electron impact ionization, by the plasma electrons around the quark nugget, is sufficient to ionize the inflowing neutrals before they penetrate significantly into the magnetopause. In the intense magnetic field within on the order of 100 times the radius of the quark nugget, the Zeeman effect provides electrons in oxygen and nitrogen atoms with potential energy that is greater than their binding energy, so those atoms will immediately ionize even at velocities too low for thermal ionization. Consequently, the background matter is sufficiently ionized when it encounters the magnetopause, and the magnetopause model is appropriate throughout the quark-nugget trajectory.

The magnetopause model lets us estimate the slowing down of the quark nugget as a function of time. For a plasma (i.e. ionized gas) with mass density $\rho_p$ and particle speed v stagnates against a magnetic field of magnitude $B$, the plasma pressure $K \rho_p v^2$ compresses the magnetic field so that the plasma pressure equals the magnetic field pressure $\frac{4B^2}{2\mu_o}$. The factor of 4 in the numerator of the magnetic pressure arises from the exclusion of the magnetic field outside the magnetopause. The plasma stagnates against the magnetic field $B = \frac{B_o r_o^3}{r_m^3}$ at the magnetopause radius $r_m$. $B_o$ and $r_o$ respectively equal the magnetic field and radius at the equator. The factor K in the plasma pressure depends on the model for the plasma; $K = 2$ for a cold plasma with elastic collisions, and $K \approx 1$ for inelastic collisions. More detailed calculations summarized by Schield[37] give $K =$ 0.832 to 0.875 for plasmas with $\gamma = 2$ and $\gamma = 5/3$, respectively, in hypersonic flow with inelastic collisions. For the purposes of our calculations of a quark nugget interacting with a wide range of ionized air, soil, and water, $K = 1$ gives sufficient accuracy. Equating these two pressures and solving for the radius $r_m$ of the magnetopause gives

$$r_m \approx \left(\frac{2B_o^2 r_o^6}{\mu_0 K \rho_p v^2}\right)^{\frac{1}{6}}. \qquad (2)$$

The corresponding cross section for momentum transfer with the magnetopause effect is $Q_m$

$$Q_m = \pi r_m^2 = \pi \left( \frac{2B_o^2 r_o^6}{\mu_0 K \rho_p v^2} \right)^{\frac{1}{3}} \quad (3)$$

The total force $F_e$ exerted by the plasma pressure on the earth is approximately

$$F_e \approx K\pi r_m^2 \rho_p v^2 \,. \quad (4)$$

Equations 3 and 4 let us calculate the energy lost during passage through matter as a function of the quark nugget mass and initial velocity. Tatsumi's order of magnitude uncertainty in $B_o$, which gives a factor of 5 uncertainty in $Q_m$ and a factor of 10 uncertainty in inferred mass. We are using Tatsumi's $B_o$ = 1 x $10^{12+/-1}$ T in the calculations in this paper and quoting the error bars for inferred mass accordingly. Obviously, a less uncertain value of $B_o$ is needed.

Results for $K$ = 1 are shown in Fig. 2 for passage of quark nuggets of mass $M_S$ and initial velocity 250 km/s[15] through the atmosphere and 3 meters of material with a density of ~1120 $kg/m^3$. The atmosphere is modeled with density increasing with altitude according to an exponential scale length of 8.5 km from 0 to 50 km. The material at sea level corresponding to water-saturated peat, serving as a witness plate for quark-nugget impacts in soil, or very salty water, serving as a target medium for the real-time detection of quark-nugget impacts.

Only velocities > 50 km/s are shown in Fig. 2 because such high velocities assure electron-impact ionization of material encountering the magnetopause occurs quickly and the magnetopause model, in its simplest form presented here, is appropriate. Quark nuggets with speeds less than hypersonic velocities in the atmosphere still ionize the surrounding matter. In the intense magnetic field within on the order of 100 times the radius of the quark-nugget core, the Zeeman effect gives atomic electrons a potential energy greater than their ionization energy. Therefore, inflowing neutral atoms quickly become ions and the stopping power remains large at less than hypersonic velocities.

As shown in Fig. 2, quark nugget velocity at the earth's surface decreases with decreasing mass, so the atmosphere provides substantial protection from low -ass quark nuggets. The threshold at which this protection occurs depends on the surface magnetic field $B_o$, as illustrated in Fig. 2.

Quark nuggets with sufficient mass to penetrate the atmosphere deposit $10^3$ to $10^6$ J energy/length, which makes them detectable.

## Detection

The enhanced energy deposition from the magnetopause effect offers at least three possibilities for detecting magnetized quark nuggets: brightness of high-velocity luminous objects in the atmosphere, acoustic waves from impacts in water, and size of craters in earth when no meteorite material was found. Each potential detection scheme is examined in this section.

*Detection by luminous events in atmosphere*

The enhanced energy deposition in the atmosphere from equation (3) lets quark nuggets be detected by the brightness and speed of the track through the atmosphere, as was investigated theoretically by De Růjula and Glashow [15]. They assumed the interaction cross section is the geometric cross section, the plasma channel behind the quark nugget is in thermal equilibrium and expands with the average molecular velocity, and radiates as a black body. They derived an expression for the luminous efficiency that is independent of the altitude and velocity of the quark nugget in the range of interest. Consequently, the luminosity is proportional to the interaction cross section. Their theory was used by Porter, *et al.*[32-33] to analyze nine data runs, collected for other purposes. The number of photons required to trigger their detector during the estimated transit time through their field of view was used to estimate the threshold luminosity of their apparatus. Then De Růjula and Glashow's theory was used to estimate the corresponding threshold for quark-nugget mass.

In some cases, the apparatus detected one event above background and the characteristics of the one event was consistent with a quark-nugget event, but the apparatus did not have two reflectors operating in coincidence mode in order to determine if the event's origin was truly astrophysical or not. Therefore, the one event was conservatively attributed to background.

Consequently, they conclude there were no "strong candidates for nuclearite [quark nugget] events." They then used Poisson statistics to give the maximum number of events consistent with this null result at the 99% confidence level and then computed the corresponding maximum flux for each observation. Fig. 3 shows their original mass assignments, assuming the interaction is the geometric cross section in equation (1), and our new mass assignments, assuming the magnetopause cross section in equation (3), for Tatsumi's predicted range of surface magnetic field $B_o$.

In order to properly interpret their results, we discussed their analysis with D. J. Fegan, one of the co-authors on Porter, *et al.*, who clarified the upper limit calculations as follows:

To derive a 99% confidence level upper limit on the flux they used the Poisson probability of obtaining $r$ events when the expected mean rate is $m$

$$P(m,r)=\frac{e^{-m}m^{-r}}{r!} \tag{5}$$

The probability of not seeing a quark-nugget event in the experiment with a background of 1 event means that r is either 0 or 1. So the total signal rate which has a 1% probability to produce either 0 or 1 recorded events is given by $e^{-m}+me^{-m}=0.01$. The solution to this equation is $m=6.6=$ signal + background. Assuming 1 event background means that the upper limit on the signal is 5.6 at 99% confidence level. Assuming no background, means the upper limit is 6.6 at 99% confidence level. They reported $m=6.1$, the average of the two calculations. The maximum flux was then computed from the observation time and area of the detector zone. This was done for all the various data runs reported by Porter, *et al*.

In essence, if the flux had been larger than indicated by the purple ellipses in Fig. 3, they would have seen events instead of a null result. If the mass had been greater than indicated for that flux, it would have still triggered their detector. Therefore, each point establishes an exclusion zone above and to the right of the data point, as is illustrated for one of the points. Lower fluxes and lower masses have not been excluded by their null results.

The solid black line is the dark-matter flux limit computed by De Růjula and Glashow for a dark-matter density of ~$10^{-21}$ kg/$m^3$ and for the expected velocity of 250 km/s *if and only if all dark matter had a single mass*. The single-mass model is relevant to subatomic dark-matter candidates[6-12] sought in high-energy particle physics experiments. However, if quark nuggets do exist, they must exist in a wide range of masses and each mass increment contributes a small fraction of the total. Consequently, the solid line in Fig. 3 is not relevant to quark-nuggets, and the exclusion zone for each particle does not exclude quark-nuggets as dark matter with a continuum of masses.

However, the combined non-excluded flux and mass of all the points in Porter, *et al.*'s analysis is much too large to be consistent with the local density of dark matter. Since the analysis does not give a mass increment with each point, more points can be added and would just make the excess worse. The problem is a dark-matter candidate with a continuum of masses requires a distribution function of masses (i.e. the flux per unit mass) such that the integral of the contributions, taken over the mass interval from 0 to infinity, provides the required dark matter density. There is no theory for the form of that distribution function, but the observations proposed below seek to provide at least a portion of that distribution function.

In addition, the above analysis implies non-magnetized quark-nuggets by assuming the interaction cross section is the geometrical cross section. The magnetopause effect increases the interaction cross section significantly and lowers the mass inferred from the luminosity in the theory by De Růjula and Glashow and the analysis by Porter, *et al.* The points from Porter, *et al.* have been reinterpreted with the magnetopause effect and are shown as the red circles in Fig. 3. There is now sufficient flexibility in the total mass to permit additional masses to contribute to the expected total for dark matter. The atmospheric observations by Porter, *et al.* may be consistent with the local dark matter density once there is enough data to construct the distribution function for quark-nugget mass. Additional observations at the Pierre Auger Observatory[38] and/or at ICE CUBE[39] could, in principle, provide sufficient data.

In addition to the paucity of data, detection in the atmosphere is complicated by the lack of experimental verification of the theory by De Růjula and Glashow, by the lack of enough photons (even with a 10-m diameter collection optic in Porter, *et al.*) to obtain a time resolved track of the luminosity and unambiguously measure the velocity, and by competing events with meteors and cosmic ray showers in the upper atmosphere. These complications may also explain why Porter, *et al.* concluded "none could be claimed as strong candidates for nuclearite [quark-nugget] events."

*Detection by impacts with water*

When a quark-nugget impacts water, the energy/length shown in Fig. 2 is sufficient to form a shock wave, which quickly decays into an acoustic pulse that travels well in water. The resulting acoustic pressure wave can be monitored with three or more, time-synchronized sensors to find the distance to the impact point by interpolation. Knowing the pressure at each sensor and the distances to the impact point, the energy per meter depth can be computed and used to estimate the mass of the quark nugget by the relationship shown in Fig. 2.

The technique is very similar to the one already used[40-42] to look for ultra-high energy (UHE) neutrinos with naval hydrophone arrays. UHE neutrinos have energy $>10^{18}$ eV. These experiments were sensitive to neutrinos depositing $> 10^{20}$ eV in the water column. Each neutrino event creates a uniquely identifiable, ~ 0.1-ms-duration, bipolar pulse. In the SAUND II observations[41], two such pulses were recorded during 130 days of observation and had energies of ~ 1 kJ and 100 kJ, which are comparable to energies expected from quark nuggets. Although the bipolar pulse was detected at the low rate of 0.015 per day, other waveforms were seen in coincidence on 3 or 4 hydrophones and passed all their other filters with a rate of 2.5/day in a 1000 $km^2$ surface area, which gives a flux of ~0.1 $km^{-2}$ $y^{-1}$ $(2\pi\text{-sr})^{-1}$ for potentially quark-nugget impacts. The corresponding quark-nugget mass from equation (5) is ~0.0004 kg. More information about the triangulated position of the acoustic source is needed to assign a tentative mass to these events.

The ACoRNE Collaboration using a more limited hydrophone array of 3 $km^2$ in area, observed two bipolar events characteristic of UHE neutrino candidates in 245 days of observation (Sloan, T., private communication, ACoRNE Collaboration, unpublished). These were selected from a total of 81 multipolar events surviving cuts against noise-generated signals. Some of these may be candidates for quark nuggets. The events were in single hydrophones without a coincidence requirement between multiple hydrophones so that the apparatus was sensitive to a much larger angular range for neutrino events than the SAUND-II experiment. The ACORNE result corresponds to a flux of 12 $km^{-2}$ $y^{-1}$ $(2\pi\text{-sr})^{-1}$. Searching the SAUND-II and ACoRNE data for pressure waveforms expected from quark-nugget impacts could efficiently and cost-effectively test the quark-nugget hypothesis.

Unfortunately, the SAUND-II and ACoRNE arrays are no longer available for scientific observations. Even if that were to change, their prime mission constrains their use and any modifications. A dedicated observatory is needed to adequately test the quark-nugget hypothesis.

Both the SAUND-II and ACoRNE arrays were located at much greater-than-optimum depth for detecting particles depositing energy near the surface. In addition, a shallow body of water with an acoustically reflective top and bottom surfaces will channel the acoustic energy, minimizing the attenuation factor. Both sets of data had significant background sounds from animals. Therefore, we sought a large body of shallow water without any sound-producing animal life and without significant human activity, but with reliable access to the cellular telephone network for communications. The southern portion of the Great Salt Lake in Utah, USA, meets these requirements. The mean depth is approximately 5 meters. The extremely high salinity prevents all animal life except algae, stromatolites, brine shrimp, and brine flies. The acoustically contiguous area is 20 km wide and 65 km long and provides a ~1300 $km^2$ detector area.

As shown in Fig. 2, we seek to detect magnetized quark nuggets with mass between 0.001 and 1.0 kg depositing 0.03 to 50 MJ/m, respectively. Three time-synchronized sensors at GPS-determined positions will record the pressure pulses and times of arrival from each impact, the location of which will be computed by solving the three, coupled, acoustic transit-time equations. Coincidence counting with the three sensors will mitigate the risk of false positives. The energy and energy/length (depth) will be computed from each pressure pulse, corrected for the attenuation during transit over the known distance between each sensor and the impact point. The three measurements will help provide an estimate of the error inherent in the process. The approximate quark-nugget mass will be inferred from the energy/length, as shown in Fig. 2.

To investigate the conversion of deposited energy in the water to pressure pulses, the CTH shock physics code[43] was used for 2D hydrodynamic simulations of a 30 MJ/m event corresponding to a 0.5 kg quark-nugget impact. In this case, the initial energy/length would be deposited within the magnetopause radius of 0.00045 m and produce an initial temperature of 2,100 eV. Radiation transport, high-temperature diffusion, and turbulent mixing from the Rayleigh-Taylor instability with the surrounding water are assumed to dominate the early dynamics of the interaction and produce a channel of larger radius and lower temperature until the non-radiative hydrodynamics will dominate the evolution of the plasma. The pulse generated by this radiation dominated phase will have a duration on the order of ten microseconds, whose high frequencies will be strongly attenuated during propagation. Therefore, the non-radiative hydrodynamics phase will dominate the pulse signature at large distances of interest.

We assume a temperature of 1 eV and an average mass density of 1000 kg/m$^3$, for the channel after the radiation-dominated phase. The corresponding radius is 0.014 m based on the SESAME[44] equation of state for water, which is used in the simulation. A full radiation-hydrodynamics treatment of this process with much higher resolution and longer simulation time will be necessary to refine the mass assigned to events, if events are indeed found.

The results of this approximate simulation for a channel extending from the surface to the full depth of 4.26 m are illustrated in Figures 4 and 5.

The cylindrical shock wave quickly decays to a cylindrical acoustic wave, shown in yellow in Fig. 4, propagating at the 1.7 km/s sound velocity. The pressure behind the pulse front is relieved by a rarefaction wave from the surface and bottom.

In addition, to the radially outward flow of water, steam and water are ejected into the air. In this 30 MJ/m case, the diameter of the evacuated region reaches about 1 m at the 4-m depth and about 2 m at the air-water interface. Air can rush into the void as the steam cools and condenses, qualitatively mitigating the amplitude of a second acoustic pulse caused by the collapse. However, mitigation of the second pulse will depend on the energy/length and will have to be studied with additional simulations that are beyond the scope of this paper.

As shown in Fig. 5, the acoustic wave changes shape as higher frequency components are preferentially absorbed, producing a sub-millisecond duration pulse. The principal results of these simulations are that a magnetized quark-nugget should produce large-amplitude, mono-polar, sub-millisecond-duration acoustic pulse that is observable at large distances on multiple platforms in accord with the transit time at the local sound speed. Although reflections from the irregular bottom and a large off-normal angle of incidence will broaden and shape the pulse somewhat, the sub-millisecond pulse shape with a fast rise and a slower fall is the signature of a quark nugget for acoustic detection in the Great Salt Lake.

The detectable event rate as a function of mass depends on 1) the incident flux as a function of mass, 2) the sensitivity of the hydrophone as a function of frequency and direction, 3) the background noise level, 4) the conversion of the deposited energy into acoustic energy, and 5) the attenuation of the acoustic pressure with distance from the impact point to the detectors. Each factor is briefly discussed, and the event rate is estimated.

As discussed in the previous section, the incident flux as a function of mass has not been established either theoretically or observationally. The null observations by Porter, *et al.*, adjusted for magnetized quark-nuggets, are of little help since they are null results and since the mass increment for each observation is unknown. Consequently, we have performed preliminary observations to see if the proposed acoustic observatory at the Great Salt Lake records enough signals for analysis. A 250-hour run with a single sensor in the summer of 2015 produced a rate of about 300 per year, which may or may not be reduced when coincidence counting with three platforms is implemented.

The signals varied from 13.5 to over 300 standard deviations of the background above background. Fig. 6 shows examples of the pressure signals.

The acoustic environment and the transmission of acoustic energy as a function of frequency had not been previously studied for the Great Salt Lake. To assess the feasibility of using it to detect quark nuggets, we conducted exploratory experiments with one sensing station located in the middle of the southern portion of the lake. The primary sensor was a C55 hydrophone [Cetacean Research, C55 series hydrophones. http://cetaceanresearch.com/hydrophones/c55-hydrophone/index.html, (2017) (Date of access: 24/06/2017)]. The sensitivity is omni-directional for horizontally propagating sound. The frequency response is -165 dB relative to 1 µPa for 30 Hz to 30 kHz and decreases to -175 dB at 50 kHz. The signal from the C55 hydrophone was passed through a 500-Hz high-pass filter to eliminate noise from the rare human activity on the Lake and then on to the signal processing electronics. The voltage signals from the pressure sensor were converted to pressure with calibration factor of 178 Pa/V. The background pressure amplitude was 92 +/- 9 mPa and the standard deviation was 3.9 +/- 1.7 mPa during the observations.

The conversion of explosive energy into acoustic energy has been extensively studied for deeply submerged spherical explosions[45] in water because of their relevance to anti-submarine warfare. However, shallow, line explosions in water have had little practical application and have yet to be studied experimentally or theoretically. In addition, the CTH simulations do not include the effects of minerals that can attenuate the higher frequencies in an acoustic pulse[42]. Therefore, we measured the pressure as a function of distance from a 112-J, 5-µs duration, wire-initiated, arc discharge extending from the surface to a depth of 1.8m. As shown in Fig. 7, the discharge produced a prompt acoustic pulse in the near field, 5-m distance from the source with ~ 10 µs duration followed by a second pulse with ~ 1 ms duration. The peak pressure pulse was reproducible within +/- 13% standard deviation and the time integral of the square of the pressure (which is proportional to the total energy in the pulse) was reproducible to +/- 26% for both the first and second pulses.

The high frequency content of the initial pulse is not satisfactorily recorded by the C55 hydrophone, so the amplitude is under reported. The high-frequency signals decay quickly with increasing distance and are overtaken by the second pulse at < 179 m from the source.

The observed pulse structure, i.e. a short-duration shock wave followed by a longer duration secondary pulse, is similar to the general behavior of underwater explosions[45] with TNT. However, the electrical discharge used here would create a small channel that is located near the surface of the lake. This small channel would not be expected to exhibit a secondary pulse due to bubble formation and collapse. The primary pulse is a direct line-of-sight measurement from the discharge, whereas the secondary pulse is a diffuse reflected wave from the bottom of the lake. At sufficient distance from the discharge the primary and secondary pulses would merge to create a unique signature of the event. Future work will address in more detail the unique signatures of an electrically driven line discharge from a calibration pulser and of the linear energy deposition expected from a quark-nugget impact.

As shown in Fig. 8, the observed pressure pulse is the net result of acoustic signals bouncing off the acoustic impedance mismatches at the air-water and water-solid interfaces. The shallow lake approximates an acoustic waveguide with constructive interference complicating the net peak pressure at any position. However, the total energy in the secondary (longer-duration) waves averages these fluctuations and is a better indicator of the total energy deposited. The total acoustic energy[46] $W$ for a cylindrically propagating acoustic pulse in a lake of constant depth $d$, with totally reflecting boundaries at the top and bottom surfaces, with water density $\rho$ and sound speed $c_s$, and at a distance $r$ from the source is

$$W = \frac{2\pi r d}{\rho c_s} \int_0^\infty P^2 dt \qquad (6)$$

Equation (6) was applied to the data shown in Fig. 8 and fit to a power law in r. The results are shown in Fig. 9.

The five points fit the power curve

$$W = \frac{0.3971}{r^{1.48}} \qquad (7)$$

with $R^2$ correlation coefficient = 0.975. Solving for $r = r_o$ at which $W = W_o = 112$ J, the total energy deposited, gives $r_o = 0.0047$ m. For the purpose of estimating the event rate, we assume $r_o = 0.0047$ m and combining equations (6) and (7) to obtain an expression for $W_o$ as a function of the time integral of the pressure squared:

$$W_o = \frac{17{,}600 r^{2.48} d}{\rho c_s} \int_0^\infty P^2 dt \qquad (8)$$

To find the distance $r$ over which the pressure is 10 standard deviations above background, i.e. $P > 0.14$ Pa, for a deposited energy $W_o$, we approximate $\int_0^\infty P^2 dt \cong P^2 (\Delta t)$ for $P = 0.14$ Pa and $\Delta t$ = the effective duration of the low frequency, acoustic pressure pulse. For the CTH simulations of the 30 MJ/m quark-nugget impact, $\Delta t = 0.4$ ms. For the 112 J deposition over 1.8 m depth in our experiments in the Great Salt Lake, the secondary pulse has $\Delta t = 1.1$ ms. Fig. 9.6 in Cole[45] shows the pressure as a function of time for the acoustic pulse produced by a 200,000 J spherical explosive charge detonated 1.8 m below the surface in seawater. From that data, we calculate $\Delta t$

= 0.89 ms. Since $\Delta t$ ~ 1 ms for these extreme cases, we conclude that $\Delta t$ ~ 1 ms for the purposes of estimating the event rate. Therefore, the threshold condition for an event is $\int_0^\infty P^2 dt \cong P^2(\Delta t) = 0.14^2(0.001) = 2\times10^{-5}$. Using that criteria and solving equation (7) for the distance r from which the sensor can record an energy deposition of $W_o$, gives

$$r_range = 248 \times W_o^{0.4}. \quad (9)$$

Since we need to communicate with each sensor to upload data and maintain the sensor, the position of the sensor system is constrained by cell phone coverage, which is excellent in the southwestern quadrant. Locating the sensor system in that quadrant and using equation (9) for the range gives the effective area for detecting quark-nuggets as a function of mass and associated energy deposition $W_o$.

As stated above, the mass distribution function of dark matter for a continuous mass distribution is not known. For the purposes of this estimate, we assume the number of events per unit mass varies inversely with mass and use the rate of ~300/year from the 250 hour single-sensor run to estimate the number of events per year near (within a factor of 3) of each quark nugget mass. The results are shown in Fig. 10.

These preliminary results indicate that detecting ms-duration acoustic signals in shallow water provides a feasible approach for testing the magnetized quark-nugget hypothesis for dark matter in the mass range of 0.001 to 1.0 kg.

Adding two more acoustic sensors and conducting radiation-hydrodynamic simulations of the generation of acoustic signals from the high-energy-density channel are obvious next steps. Then the three platforms could be used with coincidence techniques to ensure the distant (non-platform) origin of the signal. Interpolating the three observations to determine the distance to each source and to estimate the deposited energies and masses associated with the observations.

In principle, an all-sky optical sensor looking for the luminosity of the incoming track at high speed could provide a fourth and independent sensor to confirm the extra-terrestrial origin of the impacting object. However, the ambient light of nearby Salt Lake City is a complication; we have not examined the technical feasibility of an optical sensor.

*Detection by impact craters*

Rafelski, *et al.*[47] proposed that compact ultra-dense objects (CUDOs) composed of quark nuggets, with or without surrounding normal matter, could be detected by their impact craters. They considered only non-magnetized quark nuggets and concluded that the normal matter would evaporate. The remaining quark-nugget core would puncture rocky matter and penetrate deeply into rock. The crater would be distinguished from meteorite craters by the lack of meteorite and CUDOs materials. Most of the kinetic energy would continue with the quark-nugget core as it penetrates into or through the rock body. Although they focus on masses of $10^9$ to $10^{15}$ kg non-magnetized quarks, their ideas are applicable to magnetized quarks of much lower mass. With the deposition energy/length shown in Fig. 2, magnetized quark nuggets with mass > 0.001 kg should produce a substantial crater when they impact the earth.

The large energy/length suggest that such impacts should have been noticed and reported if they occur in populated areas. There have been anecdotal reports in the news that are consistent with such energetic events without meteorites. A12-m diameter crater occurred at 11:05 PM, September 6, 2014, near Managua, Nicaragua [Cooke, W. Did a meteorite cause a crater in Nicaragua? http://blogs.nasa.gov/Watch_the_Skies/2014/09/08/did-a-meteorite-cause-a-crater-in-nicaragua/ and http://www.cnn.com/2014/09/08/tech/innovation/nicaragua-meteorite/, (2014) (Date of access: 24/06/2017)]. An event occurred on July 4, 2015, at the Salty Brine Beach in Rhode Island, USA [Shapiro, E., Cathcart, C. & Donato, C. Bomb squad, ATF investigating mysterious explosion at Rhode Island beach. http://abcnews.go.com/US/explosion-report-prompts-evacuation-rhode-island-beach/story?id=32384143, (2015) (Date of access: 24/06/2017)]. Finally, an event occurred on February 6, 2016, in Tamil Nadu, India [Hauser, C. That wasn't a meteorite that killed a man in India, NASA says. http://www.nytimes.com/2016/02/10/world/asia/that-wasnt-a-meteorite-that-killed-a-man-in-india-nasa-says.html?_r=0, (2016) (Date of access: 24/06/2017)]. Each of these public events has an official investigation team dedicated to alleviating the public's concern over safety. Control of access to the event and information about the event make a scientific investigation by an outsider very difficult, if not impossible. In addition, an event rate of about 1 per year is impractical for systematic analysis.

Consequently, we searched for craters in the peat bog of County Donegal, Ireland, because peat bogs provide excellent large-area detectors for impacts because they exist over very large areas and have low yield strength. We also did computer simulations to estimate crater size as a function of energy/length deposited. The detailed results are beyond the scope of this paper and will be submitted for publication separately. In summary, peat bogs do have craters and some or all of them may have been caused by quark-nugget impacts. However, determining the quark-nugget mass from the diameter of craters is uncertain by factors of 2 to 4 and the flux from the diameter of craters is uncertain by about 50%, depending on the size of the crater. Even if experiments could accurately establish the dependence of crater diameter on line-energy deposited, there is still the possibility of unknown processes producing craters. Therefore, testing the quark-nugget hypothesis with the distribution of crater diameters is unlikely to produce unambiguous results.

## Conclusions

We have shown that high-velocity magnetized quark nuggets will interact with surrounding matter through a magnetopause, just as the earth interacts with the solar wind, and deposit about a factor of $\sim 10^6$ more energy/length than previously assumed with unmagnetized quark nuggets. The magnetopause interpretation of Porter, *et al.*'s analysis of telescope observations reduce the estimated flux as a function of mass and improve the consistency of the observations with the theoretically predicted local density of dark matter. Of three scenarios for detecting magnetized quark nuggets (by luminous tracks in the atmosphere, by impact craters in the ground, and by acoustic signals in water), we conclude that the last one is the most promising. Although reflections from an irregular bottom surface and transit time-effects from a non-normal impact angel affect the pulse shape somewhat, the signature of a quark-nugget interaction with water is a sub-millisecond duration, mono-polar, acoustic pressure pulse with a short rise time and longer duration fall time, observable at multiple distance sensors according to the local velocity of sound.

## Data Availability

All data generated or analyzed during this study are included in this published article. All the specialized software and instructions for replicating the hardware and firmware and for installing the software to make the acoustic sensors are available at primord.net.

## Acknowledgements

We gratefully acknowledge S. V. Greene for first suggesting that quark nuggets might explain the geophysical evidence that initiated this research (she generously declined to be a coauthor); W. F. Brinkman, and D. J. Fegan for an especially helpful suggestions during this work; Benjamin Hammel for conducting laboratory experiments on the interaction of a simulated quark nugget with matter; Anya Rosen-Gooding for editing the manuscript; and Jacquelyn McRae for conducting the survey of potential impact sites with Google Maps. We appreciate Mr. Josie (The Post) Duddy, Mr. and Mrs. William E. Gallagher, Rangers David Duggan and Seamus McGinty,

Emer Galligan, and Kevin Rose of County Donegal, Ireland, and David Daniels of Booze Allen Hamilton, Arlington, VA, for their assistance in the field work, and Archivists Neve Brennan and Ciara Joyce for locating survey documents. Finally, we appreciate constructive critique and suggestions from P. Cooper, B. Carr, P. Joshi, P. Steinhardt, S. Carlip, J. Malenfant, V. V. Flambaum, O'Dean P. Judd, M. Vahle, E. McGuire, K. Holley-Bockelmann, and D. Finley but acknowledging them does not connote their knowledge of, much less approval of, the final content in this paper. In addition, the team is grateful to Harbor Master David Shearer of Utah State Parks, Ms. Jamie Phillips-Barnes of Utah State Lands, Mr. David Ghizzone and Mr. Chad Wadell of Gonzo Boat Rentals, Dr. Robert Baskin of the University of Utah, Dr. Cory Angeroth of the US Geological Survey for permitting and technical assistance, and to Mr. Haydn Jones, Dr. Ed Atler, Mr. Karl Scheuch, Mr. Andrew Bloemendaal, Mr. Jonathon Cross, Mr. Red Atwood, Mr. James Bell, and Mr. Michael Rymer for technical support.

This work was supported primarily by VanDevender Enterprises, LLC, with an encouraging contribution by Dr. Karl VanDevender. The CTH simulations were supported by the New Mexico Small Business Assistance Program through Sandia National Laboratories, a multi-program laboratory operated by Sandia Corporation, a wholly owned subsidiary of Lockheed Martin Company, for the U.S. Department of Energy's National Nuclear Security Administration under contract DE-AC04-94AL85000). By policy, work performed by Sandia National Laboratories for the private sector does not constitute endorsement of any commercial product.

## Author Contributions

J.P.V. was lead physicist and principal investigator. He developed the magnetosphere model, reinterpreted Porter, *et al.,* with the magnetopause model, developed the Great Salt Lake observatory, and led the investigation on craters as an alternative detection, wrote the paper and prepared the figures, and revised the paper to incorporate the improvements from the other authors.

A. P.V. was team physicist and critiqued the evolving hypothesis to reach the final version. He also participated in the expeditions to Ireland, the collection observations, and characterization of the deformations. He reviewed and improved the paper and figures.

T. S. contributed significantly to the analysis on the atmospheric detection and the interpretation of acoustic detection in previous observations for super-high-energy neutrinos.

C. S. was lead software engineer and programmed and managed data acquisition and analysis software.

P. W. was the team geomorphologist, with a specialty in peat bogs. He participated in the field work in Ireland, developed and evaluated alternative explanations for the deformations, and decided on the interpretation of the deformations and the associated limitations. He also reviewed improved the paper and figures.

R. G. S. did the computational simulations of the shock wave from a cylindrical energy deposition in water and simulations of crater diameter versus energy for cylindrical energy deposition in water-saturated peat.

R. Z. developed the second generation, special purpose, digital signal processing firmware and software.

J.B. developed the first generation, special purpose, digital signal processing firmware and software.

J. C. contributed to the platform survivability and deployment for the Great Salt Lake observations.

N. McG. was the team historian and cultural liaison to the Irish people involved in the expeditions and participated in exploring the deformations in Ireland. In addition to providing access to the sites and the interviewed witnesses, he assured we interpreted both the deformations and the testimony in the context of the customs, practices, and language of the people. His work was essential to avoiding misinterpretation of the observations. In addition, he reviewed and improved the paper and figures.

**Additional Information**

Competing Financial Interests

The authors declare that there are no competing financial interests.

Figure Legends

**Figure 1. Magnetic field configuration of a magnetic dipole without (a) and with (b) a left-to-right flowing solar wind, illustrated as arrows, compressing the magnetic field into a magnetopause configuration.** The magnetic field is deformed by the flowing plasma until the magnetic pressure balances the plasma pressure. The magnetopause is shown as the dotted field line in (b).

**Figure 2. The velocity of 250-km/s quark nuggets after passage through the atmosphere (dashed line) and the energy per unit length (solid lines) deposited by them in passing through the first 3 m of 1120-kg/m$^3$ aqueous liquid for surface magnetic field $B_o = 10^{11}$ T (blue), $10^{12}$ T (black), and $10^{13}$ T (red).** The three values of magnetic field cover the uncertainty in Tatusmi's calculated magnetic field $B_o = 10^{12+/-1}$ T.

**Figure 3. The flux versus quark nugget mass from Porter, *et al.*[30] and the magnetopause model.** The results from Porter, *et al*. are shown in purple and are based on the geometric cross section. The masses from analysis with the magnetopause cross section are shown for Tatsumi's surface magnetic field $B_o = 10^{12+/-1}$ T in black for $10^{12}$ T, blue for $10^{11}$ T, and red for $10^{13}$ T. The width of the pattern of three ellipses indicates the dominant uncertainty in the inferred mass. The black line is the expected flux if all the dark matter had a single mass. The exclusion zone for one point shows the region in flux-mass space that the lack of observed events would exclude if and only if all of dark matter had that mass.

**Figure 4. Pressure maps at (a) 0.0 ms, (b) 1.0 ms, (c) 2.0 ms, (d) 3.0 ms, (e) 4.0 ms, (f) 5.0 ms, (g) 6.0 ms, (h) 7.0 ms, (i) 8.0 ms, and (j) 9.0 ms are shown on the right of each figure. The profile of the low-density channel and plume is shown on the left of each figure.** The air-water surface is at Z=0.

**Figure 5. Pressure as a function of time at radii (left to right in figure) of 1.2 m, 2.4 m, 3.6 m, 4.8 m, 6.0 m, 7.2 m, and 8.4 m.**

**Figure 6. Pressure versus time from single sensor observations in 2015.** The background of 94 mPa is shown as dashed line. The background plus 10 of the 4 mPa standard deviations in the background (i.e. 134 mPa) is shown as the dotted line for comparison. The background of 94 mPa pressure is the mean amplitude of the fluctuating pressure in the 500 Hz to 30 KHz band of the hydrophone, taken over 10,000 samples at 1 million samples per second. The 4 mPa is the standard deviation of those 10,000 amplitudes.

**Figure 7. (a) The average of four pulses at 5 meters from the 112-J pulser and showing the ~10-µs duration shock wave and (b) the ~1-ms duration secondary wave from the collapse of the expanding bubble.**

**Figure 8. Pressure versus time at (a) 5 m, (b) 74 m, (c) 125 m, (d) 179 m, and (e) 226 m from the pulsed source.** The time scale has been shifted in (b), (c), (d), and (e) so that the second pulses of each aligns with the second pulse at 5 m from the source. The first pulse in (a) and shown in Fig. 7 is almost too short-duration to be visible on this time scale.

**Figure 9. Energy versus distance r from the 112 J source based on equation (6) applied to the data in Fig. 8.**

**Figure 10. Estimated number of events per year near the indicated mass.**